\documentclass[11pt]{article}
\usepackage{amssymb,graphicx,latexsym}
\newtheorem{theorem}{Theorem}
\newtheorem{corollary}{Corollary}
\def\qed{$\Box$}

\newtheorem{definition}{Definition}
\newtheorem{lemma}{Lemma}
\newtheorem{observation}{Observation}
\begin{document}
\title{Constant 
communication complexity protocols for multiparty accumulative boolean 
functions} 
\author{Sudebkumar Prasant Pal\footnote{Communicating author: 
Department of Computer Science and 
Engineering and Centre for Theoretical Studies,
Indian Institute of Technology, Kharagpur 721302, India.
email: spp@cse.iitkgp.ernet.in,
http://www.facweb.iitkgp.ernet.in/\~~spp/}
\and Sima Das\footnote{Department of Computer Science and 
Engineering,
Indian Institute of Technology, Kharagpur 721302, India.}
\and Somesh Kumar\footnote{Department of Mathematics and 
Centre for Theoretical Studies,
Indian Institute of Technology, Kharagpur, 721302, India. }\\
}

\date{}
\maketitle
\begin{abstract}

Generalizing a boolean function from Cleve and Buhrman \cite{cb:sqec}, 
we consider the class of {\it accumulative boolean functions} 
of the form
$f_B(X_1,X_2,\ldots ,X_m)=\bigoplus_{i=1}^n
t_B(x_i^1x_i^2\ldots x_i^m)$, where
$X_j=(x^j_1,x^j_2,\ldots ,x^j_n), 1\leq j\leq m$ and
$t_B(x_i^1x_i^2\ldots x_i^m)=1$ for input $m$-tuples
$x_i^1x_i^2...x_i^m\in B\subseteq A\subseteq \{0,1\}^n$, and 0, if 
$x_i^1x_i^2...x_i^m\in A\setminus B$.
Here the set $A$ is the input {\it promise} set for function $f_B$.
The input vectors $X_j, 1\leq j\leq m$ are given to the $m\geq 3$ 
parties respectively, 
who communicate cbits  
in a distributed environment so that  
one of them (say Alice) comes up with the value of the function.
We algebraically characterize entanglement assisted 
LOCC protocols requiring only $m-1$ cbits of communication 
for such multipartite  
boolean functions $f_B$, for certain sets $B\subseteq \{0,1\}^n$, 
for $m\geq 3$ parties
under appropriate uniform parity promise restrictions 
on input $m$-tuples
$x_i^1x_i^2...x_i^m, 1\leq i\leq n$. We also show that these 
functions can be 
computed using $2m-3$ cbits in a purely classical deterministic setup.
In contrast, for certain
$m$-party accumulative boolean functions ($m\geq 2$), we
characterize promise sets of mixed parity for input $m$-tuples so 
that $m-1$ cbits of communication suffice in 
computing the functions
in the absence of any a priori quantum entanglement.
We compactly represent all these protocols and the 
corresponding input promise
restrictions using uniform group theoretic and hamming distance
characterizations.

\end{abstract}


\section{Introduction}
\label{introduction}

The computation of a function of several variables in a distributed environment
may require substantial communication between spatially separated
parties; typically, different components of the input are
available with the different parties, and one of the parties is
required to eventually come up with the value of the function. 
Kremer \cite{kre:qc} showed that computing the {\it two-party inner product} function 
$IP(X,Y)=(x_0y_0+x_1y_1+\ldots+x_{n-1}y_{n-1})$ mod 2, 
requires $\Omega(n)$ qubits of communication. This result holds for the 
communication complexity model given by Yao \cite{qcc:yao}, 
permiting quantum channels for communicating 
qubits between the two parties.
The linear lower bound was already known for the scenario where only
classical communication is permitted in a purely deterministic classical 
setting \cite{kre:qc,kn:cc}.
In the restricted scenario as in \cite{bcd:qecc,cb:sqec,bdht:mqcc} 
where no quantum
communication is permitted, some saving in classical
communication complexity results on exploiting
\emph{a priori quantum entanglement} and contextuality effects in
{\it quantum measurement}. 
Quantum entanglement
provides some correlation over spatially separated qubits. 
Buhrman, Cleve and van Dam \cite{bcd:qecc},
have shown that quantum entanglement can help in gaining
advantage over classical communication for certain problems.
One such problem is where three parties, Alice, Bob and Carol are each
given two-bit vectors $X=(x_1,x_0)$, $Y=(y_1,y_0)$, $Z=(z_1,z_0)$, respectively.
Alice is required to come up with 
the result of the evaluation of the function
$h(X,Y,Z)=x_1\oplus y_1\oplus z_1\oplus (x_0\vee y_0\vee z_0)$ given the
input promise $x_0\oplus y_o\oplus z_0=0$.
Buhrman et al. \cite{bcd:qecc} 
show that two cbits of communication is sufficient for Alice to
come up with the answer in the presence of three-party
a priori quantum entanglement.
This result was further used for computing $g(x,y,z)=\frac {(x+y+z) mod 4} 2$
where $x,y,z$ are two-bit integers and 
$x+y+z=0$ (mod 2). It is easy to see that $g(x,y,z)$ 
is either 0 or 1, and, is indeed the second-least significant bit in 
the binary representation of $x+y+z$.
It was shown that Alice can come up with the value of the function
with only 2 cbits of 
communication (naturally, all three parties can possess the value
after a total of 3 cbits of communication). 
The authors also established a lower bound of 4 cbits on any exact 
classical protocol generating the value of $g(x,y,z)$ at each of the parties.

A gap of one cbit between the classical lower bound and the entanglement 
assisted upper bound was also demonstrated for a three-party problem 
by Cleve and Burhman \cite{cb:sqec}. They worked on 
the three-party function
$f(X,Y,Z)=(x_1y_1z_1+x_2y_2z_2+\ldots +x_ny_nz_n)$ mod 2;
where $X$, $Y$, $Z$ are $n$ bit vectors given
to Alice, Bob and Carol, respectively. They demonstrated that
with preshared entanglement,
only two classical bits of communication is
required to compute $f$
where
the $i$th input triple $x_iy_iz_i$ is parity promise restricted 
to be of odd parity. They also showed that any classical protocol 
computing $f$ will require at least three bits of communication.
Later, Buhrman, van Dam, Hoyer and Tapp \cite{bdht:mqcc}  
considered a generalization  $F(X)$ of 
the above mentioned function $g(x,y,z)$ of Buhrman 
et al. \cite{bcd:qecc}.
This function is a partial function $F:V^m\rightarrow \{0,1\}$ where  
$V=\{0,\ldots,2^n-1\}$.
Its computation depicts a bigger 
gap (a logarithmic factor in the number 
$m$ of parties), between entanglement 
assisted communication complexity and purely classical 
communication complexity. 
This function is defined
as $F(X)=
\frac 1 {2^{n-1}}((\Sigma_{i=1}^m x_i)$ mod $2^n)$, where 
$x_i\in V=\{0,\ldots,2^n-1\}$ and 
$((\Sigma_{i=1}^m x_i)$ mod $2^{n-1}=0)$.
It is easy to observe that $F$ computes the 
$n$th least significant bit of the sum of the $x_i$'s, which is 1 if
the sum is an odd multiple of $2^{n-1}$, and 0, otherwise. 
Although the gap is 
asymptotic, a logarithmic factor in $m$, it is still a constant for 
a fixed number of parties. 
Raz \cite{exsep:raz}, demonstrated exponential communication complexity
gaps for certain partial 
functions in Yao's model \cite{qcc:yao}, where 
qubit communication is permitted. 

The most interesting results are those of 
linear lower bounds on the numbers of cbits (or qubits)
required for the two-party inner-product problem of computing 
$IP(X,Y)$ as shown by Cleve, van Dam, Nielsen and Tapp \cite{qeccipf:cdnt}, 
even in the presence of
a priori quantum entanglement. They show that such lower bounds hold for
the exact problem as well as for bounded probability of failure.  
They use a ``quantum'' reduction from a quantum information 
theory problem to the inner product problem and use a non-trivial
consequence of Holevo's theorem \cite{nc:qcqi,holevo}
to establish the lower bound.
Since quantum information subsumes classical information, this is also an
alternative proof for the linear classical communication complexity lower 
bound for the inner product problem.

\begin{table}[h!p!]
        \begin{center}
        \begin{tabular}{|c|c|c|c|c|c|c|}
        \hline
        \multicolumn{3}{|c|}{} & \multicolumn{4}{|c|}{Local operations on qubits}
\\ \multicolumn{3}{|c|}{} & \multicolumn{4}{|c|}{of $i$th entanglement }
\\ \multicolumn{3}{|c|}{} & \multicolumn{4}{|c|}{for $i$th input triple as below}\\
        \hline
        Function&Promise&Apriori entanglement&$000$&$011$&$101$&$110$\\
        \hline
        \hline
        $f_{000}$&$x_i\oplus y_i\oplus z_i = 0$& $1/2(|000\rangle-|011\rangle-|101\rangle-|110\rangle)$& $III$&$IHH$&$HIH$&$HHI$ \\
        \hline
        $f_{011}$& -do- & -do- &$IHH$&$III$&$HHI$&$HIH$ \\
        \hline
        $f_{101}$& -do- & -do- &$HIH$&$HHI$&$III$&$IHH$ \\
        \hline
        $f_{110}$& -do- & -do- &$HHI$&$HIH$&$IHH$&$III$ \\
        \hline

\end{tabular}
\end{center}
\caption{Local operations for functions $f_u$ for evaluating a single minterm $t_u$ \label{ftable}}
\end{table}
\begin{table}[h!p!]
\begin{center}
        \begin{tabular}{|c|c|c|c|c|c|c|c|c|}
        \hline

&$f_{000}$&$f_{001}$&$f_{010}$&$f_{011}$&$f_{100}$&$f_{101}$&$f_{110}$&$f_{111}$\\
\hline
$f_{000}$&&$z_i$&$y_i$&$y_i, z_i$&$x_i$&$x_i,z_i$&$x_i,y_i$&$x_i,y_i,z_i$ \\
\hline
$f_{001}$&$z_i$&&$y_i,z_i$&$y_i$&$x_i,z_i$&$x_i$&$x_i,y_i,z_i$&$x_i,y_i$\\
\hline
$f_{010}$&$y_i$&$y_i,z_i$&&$z_i$&$x_i,y_i$&$x_i,y_i,z_i$&$x_i$&$x_i,z_i$\\
\hline
$f_{011}$&$y_i,z_i$&$y_i$&$z_i$&&$x_i,y_i,z_i$&$x_i,y_i$&$x_i,z_i$&$x_i$ \\
\hline
$f_{100}$&$z_i$&$x_i,z_i$&$x_i,y_i$&$x_i,y_i,z_i$&&$z_i$&$y_i$&$y_i,z_i$\\
\hline
$f_{101}$&$x_i,z_i$&$x_i$&$x_i,y_i,z_i$&$x_i,y_i$&$z_i$&&$y_i,z_i$&$y_i$ \\
\hline
$f_{110}$&$x_i,y_i$&$x_i,y_i,z_i$&$x_i$&$x_i,z_i$&$y_i$&$y_i,z_i$&&$z_i$ \\
\hline
$f_{111}$&$x_i,y_i,z_i$&$x_i,y_i$&$x_i,z_i$&$x_i$&$y_i,z_i$&$y_i$&$z_i$&\\
\hline
\end{tabular}
\end{center}
\caption{Reducibility between functions \label{feqtable}}
\end{table}

The general 3-party partial boolean 
function may be written as a mapping from a promise restricted subset of
$\{0,1\}^n$X$\{0,1\}^n$X$\{0,1\}^n$ into $\{0,1\}$. In this paper, We consider 
3-party functions of the form 
$f(X,Y,Z)=\bigoplus_{i=1}^{n}(l_{i} \wedge m_{i} \wedge n_{i})$,
where $X=(x_1,x_2,...,x_n)$,
$Y=(y_1,y_2,...,y_n)$ and $Z=(z_1,z_2,...,z_n)$, are boolean vectors
with all the input triples
$x_iy_iz_i, 1\leq i\leq n$
obeying uniform (either even or odd) parity promise restriction.
Literals $l_{i}$, $m_{i}$, $n_{i}$ represent
$x_{i}$, $y_{i}$, $z_{i}$ appearing in the minterm
$l_{i} \wedge m_{i} \wedge n_{i}$,
either complemented or uncomplemented.
Generalizing to $m\geq 3$ parties, 
we consider the class of boolean functions 
of the form
$f_B(X_1,X_2,\ldots ,X_m)=\bigoplus_{i=1}^n
t_B(x_i^1x_i^2\ldots x_i^m)$, where
$X_j=(x^j_1,x^j_2,\ldots ,x^j_n), 1\leq j\leq m$ and
$t_B(x_i^1x_i^2\ldots x_i^m)=1$ for input $m$-tuples
$x_i^1x_i^2...x_i^m\in B\subseteq A\subseteq \{0,1\}^n$, and 0, if
$x_i^1x_i^2...x_i\in A\setminus B$.  
Here the set $A$ is the promise set for function $f_B$.
We call all such boolean functions as {\it accumulative boolean functions}.  
In all these functions, the inputs given to the $m\geq 3$ parties are $n$-bit
vectors. 
These accumulative boolean functions are inspired by
the three party function in \cite{cb:sqec}. 
For one class of eight such 3-party functions,
we algebraically characterize 
and represent protocols with preshared quantum entanglement
in Section \ref{secfoddevenparity}, where at most two 
cbits are communicated between the three parties, and 
Alice finally comes up with the value of the function (see 
Theorem \ref{flocalops}). We use the {\it Kliens four-group
(Vierergruppe)}, 
$V_4$, given by the matrix $V_4$ of Section 
\ref{secfoddevenparity}
to represent the local quantum operations necessary 
for these functions for different combinations of 
(promise restricted) inputs (see Theorem \ref{fMG}).
The promise restriction is even (or odd) 
that is, $x_i\oplus y_i\oplus z_i=0(1)$, 
where $x_i,y_i,z_i$ are
the $i$th elements of the vectors $X,Y,Z$, respectively. for $1\leq i\leq n$.   

We generalize the results of Section \ref{secfoddevenparity} 
to $m\geq 4$ parties in Section \ref{secmultihamming} for 
computing accumulative boolean functions with $m-1$ 
cbits of communication.
More precisiely, we design non-trivial promise sets and 
$m$-partite maximally entangled states using hamming 
distance characterizations for supporting entanglement assisted 
protocols (see Theorems \ref{thmhamming4} and \ref{thmhamming} and 
Corollaries \ref{corhamming4} and \ref{corhamming}). 
These $m$-party protocols use local unitary operations such as the Hadamard
operator H and the rotation operator R, as defined 
in Section \ref{secmultihamming}. 
We represent the exact manner of applying these operations for the set of all 
accumulative boolean functions considered in this paper by the 
matrices $N_m$ of Section \ref{secmultihamming}. The matrix $N_m$ 
is characterized in terms of the group represented by matrix $V_4$ 
of Section \ref{secfoddevenparity}. 
In Section \ref{secmultihamming}, we also show that $2m-3$ cbits are 
sufficient to compute these functions in the deterministic classical setting 
with no a priori quantum entanglement. The question whether the gap 
between the $m-1$ cbit entanglement assisted protocol and the $2m-3$ purely 
classical protocol can be reduced remains open.

We also study the classical communication complexity of several 
classes of 3-party functions (in Section \ref{secmixedparity}) 
in the absence of a priori quantum entanglement and 
show that two cbits of communication is sufficient for each such 
class of functions. The input promise restrictions in these cases are 
carefully chosen combinations of odd and even parities.  
In addition, we consider multiparty generalizations 
in Section \ref{secmixedparity}, 
where $m$ parties require $m-1$ cbits of communication 
but no $m$-partite a priori entanglement, for computing certain mixed parity
promise restricted accumulative boolean functions. 

The main contribution of our work is the characterization and
classification 
of various classes of accumulative boolean (partial) functions and the design
of the appropriate input promise restrictions leading to 
constant communication complexity protocols; these protocols 
typically use $O(m)$ cbits 
when $m$ parties are involved.
Use of algebraic and combinatorial 
structures and properties help us in elegantly representing 
our newly defined functions and their LOCC protocols in compact notation.
Throughout the paper we use the same commutative group $V_4$ of four 
elements and its higher cardinality generalizations as required in 
Sections \ref{secmultihamming} and \ref{secmixedparity} for 
multiparty accumulative boolean function evaluation.
Suitable a priori tripartite or multipartite quantum 
entanglements are designed for the classes of 
functions in Sections \ref{secfoddevenparity} and \ref{secmultihamming} in 
order to design $m-1$ cbit protocols; 
no quantum entanglement is needed in the case of $m-1$ cbit protocols for the 
other classes of functions in Section \ref{secmixedparity}. 



\section{Local operations for entanglement assisted protocols}
\label{secfoddevenparity}

Let $f_u$ denote the accumulative boolean function 
$f_u(X,Y,Z)=\bigoplus_{i=1}^{n}t_u(x_iy_iz_i)$
defined over input boolean vectors $X=(x_1,x_2,...,x_n)$,
$Y=(y_1,y_2,...,y_n)$ and $Z=(z_1,z_2,...,z_n)$, 
with the $i$th input triple $x_iy_iz_i$ obeying an odd or even
parity (promise) restriction.
Here, $t_u(x_iy_iz_i)=l_{i} \wedge m_{i} \wedge n_{i}$ for bit pattern $u=u_1u_2u_3$,
such that $l_i,m_i,n_i$ are $x_i(\neg x_{i}),y_i (\neg y_{i}),z_i (\neg z_{i})$,
for $u_1=1(0)$,  $u_2=1(0)$,  $u_3=1(0)$, 
respectively. We say that $t_u(x_iy_iz_i)$ is the $i$th minterm of 
type $u=u_1u_2u_3$. If $u=011$, $t_u(x_iy_iz_i)=   
\neg x_{i} \wedge y_i \wedge z_i$.
Determining $f_u(X,Y,Z)$ by computing each $t_u(x_iy_iz_i), 1\leq i\leq n$ at Alice's site would
require $n$ cbits of communication: if Bob communicates 
$y_i, 1\leq i\leq n$ to Alice, then Alice knows its own
input bit $x_i$ and can determine $z_i$ using even parity 
promise given by $x_i\oplus y_i\oplus z_i=0$. 
However, we wish to compute $f_u(X,Y,Z)$ using only 2 cbits of communication in
an entanglement assisted protocol. 

Consider the four {\it even parity} functions $f_u, u=000,011,101,110$. These 
functions are defined with input triples $x_iy_iz_i$
restricted by even parity promise set
$E^3=\{000,011,101,110\}$ for
each of the four bit patterns $u$ of even parity. (For the four odd parity 
patterns $u=001,010,100,111$, we have four more functions $f_u$, which 
we call {\it odd parity} functions. These four odd functions will have input 
triples $x_iy_iz_i$ restricted by patterns in 
the odd parity promise set $O^3=\{001,010,100,111\}$).
We develop protocols for the even parity functions; the treatment for the 
four odd parity functions is similar and symmetrical. 



In the following, we first study the (0 and 1) values of $t_u(x_iy_iz_i)$ in 
terms of $u$ and $x_iy_iz_i$, both belonging to the promise set $E$.
We then design the local operations necessary on each of the three qubits, 
for each $1\leq i\leq n$, finally, leading to the complete protocol. We need 
some notation. 
Let $u+,(x_iy_iz_i)+$ denote the successors of $u,x_iy_iz_i$, 
respectively, for values of these 3-bit patterns 
from the sequence $\langle 000,011,101,110\rangle$, where the successor of 110 roles
back cyclically to 000. We have the following observation.
\begin{observation}
\label{fucyclic}
For all $u,x_iy_iz_i$ in the sequence $\langle 000,011,101,110\rangle$, 
$t_{u+}((x_iy_iz_i)+)=t_u(x_iy_iz_i)$.
\end{observation} 
\begin{proof}
Follows from the definitions of $f_u$ and $t_u$.
\qed
\end{proof}
The value of $t_u(x_iy_iz_i), 1\leq i\leq n$, is 1 
if $u=x_iy_iz_i$, and 0, otherwise.
We first show that Alice, Bob and Carol cannot come up with bits
$a_i,b_i,c_i, 1\leq i\leq n$, 
using deterministic classical algorithms locally, such that  
$t_u(x_iy_iz_i)=a_i\oplus b_i\oplus c_i$. 
(We consider the case where $u=000$ but
other values of $u$ have similar analyses). 
If this were possible then Alice, Bob and Carol would have to 
come up (using classical deterministic algorithms) 
with boolean values $a_0(a_1)$, $b_0(b_1)$  and $c_0(c_1)$ 
depending upon $x_i, y_i$ and $z_i$ being 0(1), respectively. 
Considering the four even parity patterns
possible for $x_iy_iz_i$, we therefore have to satisfy 
(i) $a_0\oplus b_0\oplus c_0=1$,
(ii) $a_0\oplus b_1\oplus c_1=0$,
(i) $a_1\oplus b_0\oplus c_1=0$ and
(i) $a_1\oplus b_1\oplus c_0=0$.
Observe that summing up the left hand sides gives even parity whereas
we have odd parity on the right hand side, a contradiction.
We call such an impossibility as
{\it classical contextuality failure (henceforth CCF)}. This may be viewed
as a {\it non-locality game} that the three parties cannot win using any local 
deterministic classical strategy. In this game the parties can only do 
local operations but are not supposed to communicate.
Using a priori tripartite quantum entanglement however, we can 
work out local unitary operations on 
the three $i$th qubits in the three parties so that the 
resulting $i$th entanglement on (standard basis) 
local measurements gives results $a_i,b_i,c_i$ (in sites of 
Alice, Bob and Carol, respectively), such that
$t_u(x_iy_iz_i)=a_i\oplus b_i\oplus c_i$. So, the game 
can be won by the three parties using a priori entanglement and 
local unitary operations as we develop below. 
Observe that with starting
entanglement $|\psi_3\rangle=\frac 1 2 (|000\rangle-|011\rangle-|101\rangle-|110\rangle)$, 
identity operations (denoted by I) on each qubit 
keeps the entanglement unchanged, thereby leaving only even parity patterns 
of basis states on 
measurement, yielding eigenvalues +1. 
For other input triples $x_iy_iz_i\neq u$, $t_u(x_iy_iz_i)$ 
must be zero. So, we 
require to use local unitary operations on the 
three $i$th qubits in the three parties
resulting in entanglements with only odd parity 
patterns of basis states;
we note that operations 
IHH, HIH and HHI on $|\psi_3\rangle$ result in odd parity basis state patterns, 
$\frac 1 2 (|001\rangle+|010\rangle+|111\rangle-|100\rangle)$, 
$\frac 1 2 (|001\rangle+|100\rangle+|111\rangle-|010\rangle)$ and 
$\frac 1 2 (|010\rangle-|001\rangle+|100\rangle+|111\rangle)$, respectively. 
Here H denotes the one qubit Hadamard operation, given as 
$H|0\rangle=\frac {(|0\rangle+|1\rangle)}  {\sqrt 2}$ and 
$H|1\rangle=\frac {(|0\rangle-|1\rangle)}  {\sqrt 2}$. 
For $x_iy_iz_i\neq u$, the measured basis state is therefore
one of the four odd parity 
states $|001\rangle,|010\rangle,|100\rangle,|111\rangle$; the measured 
pattern of eigenvalues is used to set an even parity pattern 
$abc$ from the patterns  
110, 101, 011, 000, thereby realizing 
$t_u(x_iy_iz_i)=a_i\oplus b_i\oplus c_i=0$. 
(Basis state $|1\rangle$ has eigenvalue -1, which we 
interpret as 0, and basis state $|0\rangle$ has eigenvalue 1, intrepreted as 1). 
Symmetrically, for $x_iy_iz_i=u$, the measured basis 
state is one of the four even parity 
states $|000\rangle,|011\rangle,|101\rangle,|110\rangle$; the measured 
pattern of eigenvalues is used to set an odd parity pattern $a_ib_ic_i$ like 
111, 100, 010, 001, thereby realizing  
$t_u(x_iy_iz_i)=a_i\oplus b_i\oplus c_i=1$. 
It is now easy to assign local unitary operations corresponding to 
$t_u(x_iy_iz_i)$ as III, IHH, HIH, HHI for $x_iy_iz_i=000,011,101,110$, 
respectively, if $u=000$. Each agent can determine
whether to apply I or H to its own $i$th qubit depending
on its $i$th input bit. This gives the first row in Table 
\ref{ftable}. For the other rows we can very well 
choose the local operations to be III in the diagonal and IHH, HHI and 
HIH for $u\oplus x_iy_iz_i$ values 011, 110 and 101, respectively, 
thereby giving 
the local operations corresponding 
to $t_u(x_iy_iz_i)$ (see Observation \ref{fucyclic}). 
This completes Table \ref{ftable}. Note
that each of IHH, HHI and HIH give only odd parity basis states in the 
resulting entanglement, ensuring correct 
evaluation of $t_u(x_iy_iz_i)=0$ for $x_iy_iz_i\neq u$, as 
already explained above. 
For each $i$, the local operations generate $a_i,b_i$ and $c_i$ such that 
$t_u(x_iy_iz_i)=a_i\oplus b_i\oplus c_i$. 
Bob and Carol can communicate $b_i$ and $c_i$ to Alice 
for each $1\leq i\leq n$, 
totalling $2n$ cbits, so that Alice can compute $t_u(x_iy_iz_i)$ 
for each $i$ (and hence Alice computes $f_u(X,Y,Z)$). However, Bob (and Carol)  
may very well compute the XOR of his (her) respective 
$n$ bits $b_i$ (respectively, $c_i$), $1\leq i\leq n$, and
finally communicate just one cbit to Alice for determining $f_u(X,Y,Z)$,
totalling only 2 cbits of communication.
Now we have the entire set of protocols for each of the 
four even parity functions
$f_u$. We summarize our result in the following theorem.
\begin{theorem}
\label{flocalops}
The protocols for computing $f_u$ using only two cbits of communication
are realized using local unitary
operations $I$ and $H$ as given in Table \ref{ftable} and using $n$ sets of
a priori tripartite entanglement states $|\psi_3\rangle$.
\end{theorem}
It is not difficult to verify that 
a similar and symmetrical result holds also for odd 
parity functions $f_u$, where $u \in \{001,010,100,111\}$.

\subsection{An algebraic representation for local operations}
Now we study some algebraic properties of local operations for $f_u$, in terms
of recursively defined groups. This group theoretic study is motivated 
by the intricate but interesting patterns in 
Table \ref{ftable}. 

\begin{definition}
We define matrices $M_i$ and $M'_i$ recursively as follows.
\begin{enumerate}
\item $M_1=I$ and $M'_1=H$.
\item $M_{i+1}=\left(%
             \begin{array}{cc}
               IXM_i & HXM'_i\\
               HXM'_i & IXM_i
             \end{array}
              \right)$
\item $M'_{i+1}=\left(%
             \begin{array}{cc}
               IXM'_i & HXM_i\\
               HXM_i & IXM'_i
             \end{array}
              \right)$
\end{enumerate}
\end{definition}

In the above definition, $AXB$ denotes tensor multiplication of 
each element of the matrix $B$ by the element or entity $A$.
$M_3$ is precisely the matrix of local operations as in Table \ref{ftable} 
corresponding to terms $t_u(x_iy_iz_i)$ for functions $f_u$. 
Using bit triples $a=000,b=011,
c=101,d=110$ for III, IHH, HIH and HHI, respectively, consider the group
represented by the matrix $V_4$ below, where the rows (columns) are 
indexed from left to right (top to bottom) by group elements 
$a,b,c,d$, in that order.
The group we require for representing the local unitary 
operations for $t_u(x_iy_iz_i)$, for all $u\in \{000,011,101,110\}$ and
all $1\leq i\leq n$, 
(and therefore, for $f_u$) is given by the matrix 
$V_4=\left(%
             \begin{array}{cccc}
               a & b & c & d\\
               b & a & d & c\\
               c & d & a & b\\
               d & c & b & a
             \end{array}
              \right)$

\noindent where the $(u,x_iy_iz_i)$th element in the matrix $V_4$ is the 
element $u.x_iy_iz_i$ in the group represented by matrix $V_4$.
Here, we could imagine $a=000, b=011,c=101,d=110$ for even 
parity functions and $a=001,b=010,c=100,d=111$ for odd parity functions.
\begin{theorem}
\label{fMG}
The local operations corresponding to $t_u(x_iy_iz_i)$, 
as depicted in Table \ref{ftable} and matrix $M_3$, 
are represented by the group element
$u\odot v$ in the group represented by the matrix $V_4$, where $v=x_iy_iz_i$ 
and $\odot $ represents the group operation.  
\end{theorem}
We call the above matrix $M_3$ represented as $V_4$, the {\it game matrix} 
for the 3-party case.
Note also that each entry in $M_m$ ($M'_m$) has an even (odd) number of H
operations. We use this property in Sections \ref{secmultihamming} 
and \ref{secmixedparity}. The following lemma states a useful property of 
matrices $M_m$ and $M'_m$. This property is at the heart of the multiparty 
protocols designed in subsequent sections.
\begin{lemma}
\label{Mlocalops}
Let $m$ be the number of parties, $n$ be the size of the input bit 
vector given to each party and $u$ be an $m$-bit string of even parity.
The local operations matrix with 
entries corresponding to $u\oplus p_i$ 
indexed by $u$ in the rows and $p_i$ in the columns is identical 
to the matrix $M_m$ ($M'_m$),
where $p_i=x_i^1x_i^2\ldots x_i^m$ is the $i$th input 
$m$-tuple, $1\leq i\leq n$, of even (odd) parity. 
\end{lemma}
\begin{proof}
Proof follows by induction, using the definitions of $M_m$ and $M'_m$.
\qed
\end{proof}

\subsection{Correlation preserving reducibilities}
We now know that all functions $f_u$ can be computed with $n$ sets of a priori tripartite
entanglements and promise constrained $n$-bit vector inputs to Alice, Bob and Carol, with 
only 2 cbits of communication. In Table \ref{feqtable}, we show how we may simulate each
function in this set of eight functions by any of the other seven. The simple trick is to
toggle all bits of one or more of the three input vectors
and accordingly choose the simulating function; the promise automatically 
gets set as required in the simulations. (When bits of an odd number of vectors are toggled,
the parity must switch). This equivalence also implies (following the
lower bound proof in Cleve et al. \cite{cb:sqec}), that each of these eight functions 
has a classical computation protocol with 3 cbits of communication. 
In addition, 
this equivalence also implies that none of these 
functions can be computed using 2 cbits 
of communication. We summarize these facts in the following theorem.

\begin{theorem}
Each of the eight functions $f_u$ can be computed by a classical protocol that requires only 
three cbits of communication. Moreover, none of these functions has a two cbits 
classical communication protocol. 
\end{theorem} 

The above simulation of one function by any of the seven other functions
is done using reductions that do not alter correlations between
bit vectors given to the three parties. We call such reductions as {\it
correlation preserving reductions}.

\section{Hamming distance characterizations of promise 
sets}
\label{secmultihamming}

In this section we extend entanglement assisted protocols
requiring constant classical communication complexity, to accumulative
boolean functions for $m\geq 4$ parties. Extending the protocols of
Section \ref{secfoddevenparity} essentially means spelling out
local operations in each of the $m$ parties; we do this by using the 
matrix $M_m$ of Section \ref{secfoddevenparity}. 
We state the required definitions 
and notation. Let $E^m$ ($O^m$) denote the set of $2^{m-1}$ even (odd) 
parity $m$-bit strings.   
We denote the (even parity) functions as $f_u(X_1,X_2,...,X_m)=
\bigoplus_{i=1}^{n} t_u(x_i^1x_i^2...x_i^m)$, where $u\in E^m$, and
$t_u (x_i^1x_i^2...x_i^m)$ is 1 for $x_i^1x_i^2...x_i^m=u\in A\subseteq E^m$, 
and 0, otherwise.
(A similar and symmetric definition is possible for 
odd parity functions). Here, the set $A$
is the input promise set to which the input bit strings 
$x_i^1x_i^2...x_i^m,1\leq i\leq n$, are restricted.
We characterize certain promise subsets $A\subseteq E^m$, 
permitting entanglement 
assisted protocols using exactly $m-1$ cbits of communication, using
$n$ sets of $m$-partite maximally entangled states, and
local unitary operations governed by matrix $M_m$. 
For $m=4$, we show that the 
permissible promise sets are 
$A\subseteq E^4\setminus \{x |$ where $(u\oplus x)=1111, x\in E^4\}$.  
So, for $f_{0001}(X_1,X_2,X_3,X_4)$, the promise sets that work are 
$A\subseteq E^4\setminus \{1110\}$, with a unique entangled state that 
we develop below; this
entangled state contains the eight odd parity basis states. 
Finally, we also consider cases where $m\geq 5$.
For these generalized mutiparty cases, we define accumulative boolean functions
$f_B(X_1,X_2,\ldots ,X_m)=\bigoplus_{i=1}^n 
t_B(x_i^1x_i^2\ldots x_i^m)$, where
we define
$t_B(x_i^1x_i^2\ldots x_i^m)=1$ for input $m$-tuples 
$x_i^1x_i^2...x_i^m\in B\subseteq A\subseteq O^m$, and 0, otherwise.
Here the set $A$ is the promise set for function $f_B$.

\subsection{Promise sets and entangled states}

Restricting $m$-party local operations to those 
defined by matrices $M_m$, we first 
establish a few results correlating choices of superposition patterns that use 
all the odd (or even) parity basis states
in maximal $m$-partite entanglement states. 
In particular, we would be considering local operations as 
given in $M_m$ and entanglement state
$|\psi_m\rangle={\frac 1  {2^{(m-1)/2}}} 
\sum_{v\in O^m} (-1)^{sg(v)} |v\rangle$, where 
$sg(v)=1$ only for superposition
basis states carrying minus sign, and $sg(v)=0$, otherwise. 
We derive a suitable functions $sg$ for our protocols below.
We need some notation. Let $|s=s_1s_2...s_m\rangle$ denote 
an $m$-partite standard basis state in $O^m$ 
in the $2^m$-dimensional Hilbert space $H^{\otimes m}$. Let $|s_{ij}\rangle, 1\leq i< j\leq m$, denote the (sub)state of
$|s\rangle$ in the $2^{m-2}$-dimensional Hilbert space 
$H^{\otimes(m-2)}$ with the qubits of the $i$th and the $j$th 
parties in $|\psi_m\rangle$ dropped.
We use the notation $H_iH_j$ to denote the operator where local 
Hadamard operations are performed on the $i$th and $j$th qubits in the 
respective sites and the identity operation is performed on all other qubits.
First we establish the following result.
\begin{lemma}
\label{lemma2H}
Given two basis states $|s\rangle$ and $|t\rangle$ in $O^m$, 
separated by hamming distance two, 
let $|s_{ij}\rangle=|t_{ij}\rangle$, for some
$1\leq i<j \leq m$. Then, $H_i\otimes H_j |\psi_m\rangle$ will get 
only even parity $m$-partite basis state superpositions if 
we set $sg(s)$ and $sg(t)$ such that 
$s_i\oplus s_j\oplus sg(s)\oplus sg(t)=1$.  
\end{lemma}
\begin{proof}
It is easy to see that $H^{\otimes 2} |01\rangle$ and 
$H^{\otimes 2} |10\rangle$ have $|01\rangle$ 
and $|10\rangle$ states with opposite signs. Moreover, 
$|s\rangle$ and $|t\rangle$ 
have hamming distance two, with bit disagreement only at the $i$th and $j$th
positions. 
So, in case (i) if $s_i\oplus s_j=1$ 
(and therefore $|s_{ij}\rangle$ 
and $|t_{ij}\rangle$ have
even parity), 
we assign identical signs $sg(s)=sg(t)$. Likewise, 
in case (ii) if $s_i\oplus s_j=0$
(and therefore $|s_{ij}\rangle$ 
and $|t_{ij}\rangle$ have
odd parity), 
we assign opposite signs $sg(s)$ and $sg(t)=1\oplus sg(s)$. 
Such assignments for function $sg$ would ensure cancellation of 
all odd parity basis states.
\qed
\end{proof}

For instance, consider $f_{0001}$ (without loss of generality).
Consider basis states $a=|0001\rangle$ and $|b=0010\rangle$ superimposed in 
the shared a priori entangled state $|\psi_m\rangle$, where $m=4$. Considering 
input quadruple $0001$, the entanglement remains unchanged
due to operations $I^{\otimes 4}$; 
so, standard basis measurements at the four sites
will result in odd parity basis state patterns. Whereas
for input 
quadruple $0010$, matrix $M_4$ shows that we need to do H operations
on the third and fourth qubits and no operations on the first two qubits.
The IIHH operation on basis states 
$a=|0001\rangle$ and $b=|0010\rangle$ will lead to 
cancellation of all odd parity basis states $|0010\rangle$ and $|0001\rangle$
if $a$ and $b$ have the same probability amplitude with identical +/- signs
(as stated in Lemma \ref{lemma2H}).
Considering the same input quadruple $0010$ again, and 
applying Lemma \ref{lemma2H}, 
we see that we must also give same signs for the pair of 
basis states
$(g=|1101\rangle,h=|1110\rangle)$, but different signs for the pairs 
$(c=|0100\rangle,d=|0111\rangle)$, and $(e=|1000\rangle,f=|1011\rangle)$. Assigning such signs will 
ensure that the resulting 4-partite entangled state will give 
odd parity basis
states on standard basis measurements at four sites.
Similarly, 
considering five more 4-bit input quadruples $0100,0111,1000,1011$ and $1101$, 
we can deduce applying Lemma \ref{lemma2H} that basis states' pairs which must
agree on their signs are respectively, $(a,c)$ and $(f,h)$, 
$(b,c)$ and $(f,g)$, 
$(a,e)$ and $(d,h)$, $(b,e)$ and $(d,g)$, and 
finally, $(c,e)$ and $(d,f)$, whereas,
basis states' pairs which must
disagree on their signs are respectively, $(b,d)$ and $(e,g)$, 
$(a,d)$ and $(e,h)$, $(b,f)$ and $(c,g)$, and finally, $(a,g)$ and $(b,h)$.
With some thought, it follows that the unique 
solution is to assign the same sign to basis
states $a,b,c,e$ and just the opposite sign to 
basis states $d,f,g,h$. Since we have considered 
the function $f_{0001}$, the input quadruples considered were
in the promise set $O^4\setminus \{1110\}$. Generalizing over all $u\in 
O^4$, we can now state the following results, where $u\oplus u'=1111$.
\begin{theorem}
\label{thmhamming4} 
Let $u\in O^4$ and $u\oplus u'=1111$. 
Let the input quadruples $x_i^1x_i^2x_i^3x_i^4, 1\leq i\leq n$ be 
restricted to 
elements of any promise set $A\subseteq O^4\setminus \{u'\}$. Using
$n$ instances of the entangled 
state $|\psi_4\rangle$, and local 
operations as in matrix $M_4$, it is possible   
for Alice to come up with 
the value of $f_{u}(X_1,X_2,X_3,X_4)= 
\bigoplus_{i=1}^{n} t_u(x_i^1x_i^2x_i^3x_i^4)$, with only three cbits
of communication.
\end{theorem}


\begin{corollary}
\label{corhamming4}
Let $u\in O^4$ and $u\oplus u'=1111$.
Let the input quandruples $x_i^1x_i^2x_i^3x_i^4, 1\leq i\leq n$ be 
restricted to 
elements of any promise set $A\subseteq O^4$. Using
$n$ instances of entangled state $|\psi_4\rangle$, and 
local operations as in matrix $M_4$, it is possible for  
Alice to come up with the value of
$f_{\{u,u'\}}(X_1,X_2,X_3,X_4)= 
\bigoplus_{i=1}^{n} (t_u(x_i^1x_i^2x_i^3x_i^4)\oplus 
t_{u'}(x_i^1x_i^2x_i^3x_i^4))$, for any $u\in O^4$, with only three cbits of
communication.
\end{corollary}
\begin{proof}
It turns out that $H^{\otimes 4}$ operating on $|\psi_4\rangle$ (for input
$u'$) yields an 
entanglement state with only odd parity basis states. The effect is 
same as that with operations $I^{\otimes 4}$ for input $u$. So, clubbing $u$ 
and $u'$ together for $f_{u,u'}$ using minterms 
$t_u$ and $t_{u'}$ does the needful.
\qed
\end{proof}

It is interesting to note that 
we chose 
to assign plus and minus signs in such a manner to the basis states 
in the maximal entanglement $|\psi_4\rangle$ 
that the basis states with the
same number of 1's 
got identical signs. 
This also holds for the tripartite 
entanglement $\frac 1 2 (|001\rangle+|010\rangle+|100\rangle-|111\rangle)$ 
used by Cleve et al. 
\cite{cb:sqec}, 
in their entanglement assisted protocol for
computing $f_{111}(X,Y,Z)=\bigoplus_{i=1}^{n}x_i\wedge y_i\wedge z_i$,
with only 2 cbits of communication and odd parity
promise over input triples $x_iy_iz_i,1\leq i\leq n$.

\subsection{Promise sets for the general case of multiple parties}

For the multiparty accumulative boolean functions, we 
now pose the general version 
of the {\it non-locality game}, whose 3-party version was analyzed in Section 
\ref{secfoddevenparity}. In this game, we require the $j$th of the 
$m$ parties to receive its 
respective input bit $x_i^j$ and come up with boolean value 
$a_i^j$ such that $t_u(x_i^1x_i^2...x_i^m)=
a_i^1\oplus a_i^2...\oplus a_i^m$. 
This is not possible in 
a purely deterministic classical setup but possible when a priori 
multipartite entanglement is used. Note that the parties cannot 
communicate in this 
game but may perform local operations.

So far we considered using only local operations I and H in our 
protocols. We now consider use of operators H and a rotation operator
R defined as R$|0\rangle=|0\rangle$ 
and R$|1\rangle=e^{\frac {i\pi} 2} |1\rangle$.
Let $N_m$ be the matrix obtained from matrix $M_m$ by replacing (i) I with
H and (ii) H with HR. 
Let $|\psi_m^{GHZ}\rangle= \frac {|0^m\rangle+|1^m\rangle} {\sqrt 2}$ 
be the maximally entangled $m$-partite GHZ state (also called the 
$m$-CAT state).  
We establish the following results.

\begin{lemma}
\label{mgame}
Let the $i$th input triple be $v=x_i^1x_i^2...x_i^m$, where the $j$th party is
given bit $x_i^j$, $1\leq j\leq m$. 
Suppose the $j$th party, $1\leq j\leq m$, performs an H (HR) operation 
provided $x_i^j$ is equal (not equal) to the $j$th bit 
of $u$. Then, it is possible for 
the $j$th party to come up with bit $a_i^j$ such that 
$t_u(x_i^1,x_i^2...x_i^m)=a_i^1\oplus a_i^2...\oplus a_i^m$. 
\end{lemma}
\begin{proof}
If the $i$th input triple $v=x_i^1x_i^2...x_i^m$ is identical to 
$u$, we simply 
perform $H^m|\psi_m^{GHZ}\rangle$, giving only even parity 
basis states in the resulting entanglement. For $v\in O^m$ such that 
$v$ and $u$ have hamming distance equal to an odd multiple (say $2k$, where
$k$ is odd) of 2, we observe
that local operation HR is performed at $2k$ locations. This results in a
local phase factor of $(e^{\frac {i\pi} 2})^{2k}=e^{ki\pi}=-1$ for the 
second term in $|\psi_m^{GHZ}\rangle$,
flipping its sign. With the H operations now at all the $m$ sites,
the resulting entangled state has only the
odd parity basis states. So, after performing standard basis 
measurements at the $m$ sites, 
the measured values of local qubits can be represented at their respective 
sites as boolean values $a_i^j$, such that 
$a_i^1\oplus a_i^2...\oplus a_i^m$ is of odd parity if and only if $v=u$. 
Thus, we have $a_i^j$ generated at the $j$th site 
such that $t_u(x_i^1,x_i^2...x_i^m)=a_i^1\oplus a_i^2...\oplus a_i^m$, 
winning the non-locality game. 
\hfill \qed
\end{proof}

\begin{theorem}
\label{thmhamming} 
Let $u\in O^m$. 
Let the input $m$-tuples 
$x_i^1x_i^2\ldots x_i^m, 1\leq i\leq n$, be 
restricted to the elements of 
any promise set $A\subseteq \{v | v\in O^m,$ and either $v=u$ or 
$v\oplus u$
has parity equal to an odd multiple of 2$\}$. Using
$n$ instances of entangled state $|\psi_m^{GHZ}\rangle$, 
and local operations as in matrix $N_m$, it is possible   
for Alice to come up with the value of 
$f_{u}(X_1,X_2,\ldots ,X_m)= 
\bigoplus_{i=1}^{n} t_u(x_i^1x_i^2\ldots x_i^m)$, with only $m-1$ cbits
of communication.
\end{theorem}
\begin{proof}
Computing $f_{u}(X_1,X_2,\ldots ,X_m)$ requires evaluating 
the XOR of terms
$t_u(x_i^1x_i^2\ldots x_i^m)$, 
where each term
can be written as 
$a_i^1\oplus a_i^2...\oplus a_i^m$, 
$1\leq i\leq n$, as shown in Lemma \ref{mgame}. We can compute $A_j$, the XOR of 
$a_i^j$, $1\leq i\leq n$ in the $j$th party locally, for each $1\leq j\leq m$. Then,
using $m-1$ cbits of comunication,
the bits $A_j$, $2\leq j\leq m$, can be communicated to the first party for evaluation of
$f_{u}(X_1,X_2,\ldots ,X_m)$. 
\end{proof} \hfill \qed

\begin{corollary}
\label{corhamming} 
Let $u\in O^m$. 
Let $B\subseteq O^m$ be any set of elements $v\in O^m$ (including $u$),
such that 
$u$ and $v$ have hamming distance equal to an even multiple of 2. 
Let the input $m$-tuples 
$x_i^1x_i^2\ldots x_i^m, 1\leq i\leq n$ be 
restricted to the elements of 
any promise set $A\subseteq O^m$.  
Using
$n$ instances of entangled state $|\psi_m^{GHZ}\rangle$, 
and local operations as in matrix $N_m$, it is possible   
for Alice to come up with the value of 
$f_{B}(X_1,X_2,\ldots ,X_m)= 
\bigoplus_{i=1}^{n} t_B(x_i^1x_i^2\ldots x_i^m)$, with only $m-1$ cbits
of communication.
\end{corollary}
\begin{proof} The promise set $A\subseteq O^m$ is arbitrary here. For 
elements of $B$, note that 
the phase term of $(e^{\frac {i\pi} 2})^{2k}=e^{ki\pi}=+1$ for the
second term in $|\psi_m^{GHZ}\rangle$, 
leaving its sign intact because $k$ is even. 
With the H operations now at the $m$ sites,
the resulting entangled state has only the
even parity basis states, like what happens when we apply $I^{\otimes m}$
to $|\psi_{GHZ}\rangle $. So, we can club $u$ and the entire set $B$ together
as distinguished from the rest of the elements of $O^m$. Hence we can compute
$f_B$ using $m-1$ cbits of communication by correctly determining the 
parity  
$t_B(x_i^1x_i^2\ldots x_i^m)$, $1\leq i\leq n$.
\qed
\end{proof}

We end this section with a classical protocol scheme for computing 
such promise restricted 
functions by observing that all the functions $f_u, u\in \{0,1\}^m$,
are mutually reducible as depicted in Table \ref{ftable}. 
So, it suffices to deal 
with $f_{0^m}$. The promise set comprises even parity $m$-tuple patterns that  
have 2, 6, ..., 2(2k-1), ...  1's 
in the pattern, 
contributing 0's to the function, and the 
pattern $0^m$ contributing 1. Alice therefore
needs to determine $(n-p)$ mod 2, where $p$ is  
number of non-zero $m$-tuples; it is easy to see that $p$ is half
the modulo 4 sum of the total number of non-zero bits given as inputs to
the $m$ parties. So, the $m-1$ parties can compute the modulo 4 sum of 
non-zero bits in their respective input vectors and pass on the two bits 
to Alice, whence she can compute $p$ and $(n-p)$ mod 2 as the value of the 
function $f_{0^m}$.
This results in classical communication complexity $2m-2$ cbits, which can 
further be reduced by one bit 
since we know that input $m$-tuples are of even parity.

\section{Classical protocols for accumulative boolean functions with 
mixed parity promise}
\label{secmixedparity}

Unlike functions $f_u$ where the promise was strictly based on 
either even or odd parity, we 
now consider new classes of functions where input triples are restricted 
by various mixed 
parity constraints. We characterize (i) these promise sets, and (ii) the 
LOCC protocols for computing these accumulative boolean 
functions with (a constant 
number of) $m-1$ cbits where $m\geq 3$ is the number of parties.   
Let $g_u(X,Y,Z)=\bigoplus_{i=1}^{n}t_u(x_iy_iz_i)$. Here, as in the 
case of $f_u$, $t_u(x_iy_iz_i)$ is 
again a minterm determined by the bit pattern $u$. 
First consider $u=000$ where we restrict
input $x_iy_iz_i$ to the elements of the 
set $O=\{000,001,010,100,111\}$. Note 
that in this case we have a mix of even and odd parities, 
with 000 coming along with 
all the four odd parity patterns. With the same four odd 
parity patterns, we can define three more functions $g_u$, 
where $u=011,101,110$; the
input patterns in the promise 
set being $\{u,001,010,101,111\}$. We reiterate that
$t_u(i)=1$ if and only if $u=x_iy_iz_i$, very much as 
in the case of functions $f_u$. 

\subsection{Protocols for inputs with mixed parity promise}
Now we follow the design technique similar to the one in the previous sections
for coming up with protocols for Alice computing $g_u$ for input vectors $X,Y,Z$ given
to Alice, Bob and Carol, respectively, obeying promise restrictions as just mentioned.
Note that the pattern $u$ has even parity. So, for $t_u(u)$, we may very well settle 
with an even number (may be none) of toggling local operations over $x_iy_iz_i=u$, 
keeping the inter-party
parity over the input $x_iy_iz_i$ unchanged even after toggling. This is 
indeed possible if we start with a bit pattern $S_i=u$ for the $i$th triple, 
where one bit of $S_i$ is in each party, and we toggle the respective bits 
in each party if the XOR of the 
input triple bit $x_i$, $y_i$ or $z_i$, for the respective party, 
with the respective bit in 
$u$ is 1, and, do nothing otherwise. 
Since, $t_u(x_iy_iz_i)=0$ for $x_iy_iz_i\neq u$, 
we require to get a zero contribution in such cases; 
due to odd hamming distance between the promise permitted odd input
parity triples 
$x_iy_iz_i\neq u$, and the even parity of $u$, only an odd number of 
toggling operations can result in toggling operations controlled by the odd parity pattern
$u\oplus x_iy_iz_i$.
Since we start with even parity $S_i=u$, this action will result in $S_i$ 
gaining odd parity only for input triples $x_iy_iz_i$ of odd parity. This holds
for any even 
parity pattern $u$ (and, therefore for all functions $g_u$ for even parity patterns $u$). 
Once this step is over, we observe that if the $i$th input triple is $u$, then $S_i$ 
will result in an even parity patterns; otherwise, $S_i$ will end up with odd parity. 
Naturally, toggling all bits in all $S_i$ now will result in odd parity patterns 
for input triple $u$ and even parity for others. Indeed, all we need 
to do at this stage is to compute XOR of all $3n$ bits 
of $S_i, 1\leq i\leq n$, yielding $g_u(X,Y,Z)$. 
The rest of the protocol is identical to the remaining steps 
of protocols for $f_u$. In particular, local XOR over three $n$-bit
vectors is used before the two parties Bob and Carol communicate one bit each to Alice.
The only difference is that we use pattern $S_i=u$, which is a 
classical state (say, 000 for $u=000$), and our local operations were simply toggling
classically 0 and 1 states. We summarize this fact in the following theorem; 
we also generalize this result in Section \ref{subsecmultiminterm}  
to an $m-1$ cbits classical protocol for the 
$m$-party versions of such mixed parity functions using the 
group $M_{m}$ and $M'_{m}$ of Section \ref{secfoddevenparity}.  

\begin{theorem}
\label{theoremg}
The eight three-party functions $g_u$ of mixed parity promise 
set $O^3\cup \{u\}$ can be computed 
where Alice comes up with the value of the function, using only two cbit
distributed protocols requiring no a priori quantum entanglement and only local 
classical operations. 
\end{theorem}

\subsection{Extension to multiminterm functions}
\label{subsecmultiminterm}

Extending the above ideas, note that we may increase the cardinality of the 
promise sets by adding other even parity patterns in addition to $u$ for $g_u$, giving 
rise to new functions say $g_{uu'}$, where only input triples 
$x_iy_iz_i$ of even parity result in 
the value 1 for $t_{uu'}(x_iy_iz_i)$, and therefore the 
$i$th term $t_{uu'}(x_i,y_i,z_i)$ may be viewed as a multiminterm
boolean expression with XOR (or OR) operation between them. For instance,
with $u=000$ and $u'=101$, we have $g_{uu'}(X,Y,Z)=
\bigoplus_{i=1}^{n}
((\neg x_i\wedge \neg y_i\wedge \neg z_i) \oplus 
(x_i\wedge \neg y_i\wedge  z_i))$.
We can have six choices of $u,u'$, combinations 
without repetitions of four patterns from the set $E$ taken two at a time. 
Likewise, for three minterms, 
we will have four functions, and only one function if we take all four even parity minterms. Result
similar to Theorem \ref{theoremg} holds for all these functions.
In summary, the promise sets for these functions are $A\cup O^3$, where 
$A\subseteq E^3$, and $O^3$ and $E^3$ are 
the sets of three bit
odd and even parity patterns, respectively, as defined in Section \ref{secfoddevenparity}.
The protocol remains similar to the one corresponding to Theorem \ref{theoremg} for the computation of 
functions $g_u$. We summarize the result as a corollary.

\begin{corollary}
\label{corollarygA} 
Consider three party accumulative boolean
functions $g_A(X,Y,Z)=\bigoplus_{i=1}^{n} t_A(x_iy_iz_i)$, 
where (i) $A\subseteq E^3$, (ii) the input promise restricts 
$x_iy_iz_i\in A\cup O^3$, and (iii) $t_A(x_iy_iz_i)$ is
1 for $x_iy_iz_i\in A$, and 0, otherwise.
The functions $g_A$ of such mixed parity promise can be computed with Alice coming up 
with the function value with only two cbit
distributed protocols requiring no a priori quantum entanglement and only local 
classical operations. 
\end{corollary}


Generalization of the basic result in Theorem \ref{theoremg} to multiple 
parties is as follows. The function $g_A(X_1,X_2,...,X_m)=
\bigoplus_{i=1}^{n} t_A(x_i^1x_i^2...x_i^m)$, where  
$t_A(x_i^1x_i^2...x_i^m)$ is 1 for  
$x_i^1x_i^2...x_i^m\in A\subseteq E^m$, and 0, otherwise. The input promise
set is $A\cup O^m$ as in Theorem \ref{corollarygA}.

For odd number $m\geq 5$, we start with any even 
parity bit 
vectors $S_i=s_i^1s_i^2...s_i^m,1\leq i\leq n$, where bit $s_i^j$ is 
locally held by the $j$th party. For function $g_A$, where $A\subseteq E^m$, 
the $j$th
party toggles $s_i^j$ if and only 
if the $j$th bit in $u\oplus x_i^1x_i^2...x_i^m$
is 1. Here, the first party is Alice and the input vector to the $j$th party is 
$X_j=x_1^jx_2^j...x_n^j$.
Clearly, $S_i$ attains odd parity only for odd parity input $m$-tuples 
$x_i^1x_i^2...x_i^m$. Since $m$ is odd, we can toggle all bits of $S_i$
locally to get even parity $S_i$ for even parity input $m$-tuples
$x_i^1x_i^2...x_i^m$. 
For computing $g_A(X_1,X_2,...,X_m)$, we now need to perform 
XOR over all bits of all $S_i,1\leq i\leq n$. This can be achieved by doing
XOR universally over all $s_j^i$ locally at the $j$th site 
for $i\leq i\leq n$, and then communicating these
results to Alice from each party $j,2\leq j\leq m$, 
using a total of $m$ cbits. Alice can then do the obvious rest.

When $m\geq 4$ is even, all we need to do is choose odd parity 
$S_i, 1\leq i\leq n$, to begin with. 
It is easy to now check that 
the rest of the steps are similar to the case where $m$ is odd, and we also do not need
the final universal toggling step over all bits of all $S_i$.
The result is summarized as follows.

\begin{theorem}
\label{theoremgAm}
Consider the multipartite accumulative boolean functions 
$g_A(X_1,X_2,...,X_m)=\\ \bigoplus_{i=1}^{n} t_A(x_i^1x_i^2...x_i^m)$ 
with $j$th party getting 
input bit vector $X_j, 1\leq j\leq m$, where  
$t_A(x_i^1x_i^2...x_i^m)$ is 1 for  
$x_i^1x_i^2...x_i^m\in A\subseteq E^m$, and 0, otherwise. 
The input promise restricts 
$x_i^1x_i^2...x_i^m\in A\cup O^m$. 
The multiparty functions $g_A$ of such mixed parity promise can be 
computed with Alice coming up 
with the function value with only $m-1$ cbit
distributed protocols requiring no a priori quantum entanglement and only local 
classical operations. 
\end{theorem}

\section{Concluding remarks}

We investigated different types of promise sets for input 
tuples for evaluating accumulative boolean functions with constant 
communication complexity. We demonstrated purely classical $m$-party protocols
requiring $m-1$ cbits of communication for mixed parity promise sets
where tuples of opposing parities (even and odd) 
contribute 1's and 0's respectively, to 
the accumulative boolean function (see 
Section \ref{secmixedparity}). Here, one or more $m$-tuples 
of even (odd) parity may be permitted, contributing 1's, 
just as multiple non-contributing $m$-tuples 
of the opposite parity are permitted as inputs, contributing 0's. 
For input promise sets containing only even (or odd) parity
bit patterns, we designed
constant communication complexity 
entanglement assisted protocols for such accumulative boolean functions
with $m-1$ cbits for the $m$-party case 
(see Section \ref{secfoddevenparity} and \ref{secmultihamming}). Here,
discrimination is made between a specific even (or odd) parity $m$-bit 
input string (or a suitably defined subset of input strings)
against all the input strings from a specific promise subset of the 
remaining $m$-bit input strings of the same parity. 
We designed the requisite maximally entangled states using the eight basis 
states of odd parity, by assigning real 
probability amplitudes of equal magnitudes to the basis states for the 4-party
case. The signs of these amplitudes had to be chosen carefully 
in accordance with the chosen promise sets,
as characterized in Section \ref{secmultihamming}.
For the general $m$-party problem, $m\geq 3$, we designed an alternative 
entanglement assisted protocol
in Section \ref{secmultihamming},
using $m-1$ cbits of communication and $n$ copies of the $m$-CAT entangled 
state. 

We are currently investigating along similar lines, looking for more such 
algebraic structures and characterizations.  
Swain \cite {s:meaedc} reports constant communication complexity
protocols for a class of accumulative $m$-party 
boolean functions that compute {\it 
total disagreement parity}, of (say) Alice with the rest of the parties
over multiple $m$-tuples.
The promise sets are suitably defined in a different manner; the
a priori entanglements and local operation matrices too are different 
from what we use in this paper. 

It is worthwhile unifying the protocols in this paper in terms of the
patterns of local operations performed in each of the parties; it is 
indeed possible to find a common line in all 
the protocols in this paper 
based on the group represented by matrix $V_4$. Consider Theorems
\ref{fMG}, \ref{thmhamming4}, \ref{thmhamming}, 
\ref{theoremg} and \ref{theoremgAm}, and Corollaries \ref{corhamming4},
\ref{corhamming} and \ref{corollarygA}. 
Observe the functions $g_u$ and $g_A$ 
in Section \ref{secmixedparity}, where 
a non-trivial local (toggling) 
operation is done on the $j$th bit of $S_i$ based on 
the $j$th bit in the pattern $u\oplus p_i$, where 
$p_i=x_i^1x_i^2...x_i^m$ is the $i$th input $m$-tuple, $1\leq i\leq n$. 
If this bit is 0, the identity operation I is done. If this bit is 1,
the toggling operation is 
done. 
The local operations comprise an
even number of toggling operations
corresponding to $u\oplus p_i$ 
indexed by $u$ in the rows and $p_i$ in the columns, where 
$u$ as well as $p_i$ are of even parity.
Symmetrically,  
local operations corresponding to $u\oplus p_i$ 
has an odd number of toggling opearations, where $u$ has 
even parity and $p_i$ has odd parity. In the case of 
functions $g_u$ and $g_A$, if $p_i$ is even, the even parity 
in $S_i$ is not disturbed by any local operation pattern, whereas, 
for odd parity $p_i$, the parity in $S_i$ must be reversed. In this manner the 
protocols in Section \ref{secmixedparity} correctly compute the partial
functions $g_u$ and $g_A$ with mixed parity promise sets.
It is easy to see that the matrix $M_m$ ($M'_m$) compactly encodes the XOR
operation in its entry 
corresponding to the pattern $u\oplus p_i$ in the $u$th row and $p_i$th column, if $p_i$ and $u$ agree (differ) in parity.
(The matrix entry must be translated replacing I by 0 and 
H by 1). In Section \ref{secmultihamming} we used matrix $N_m$ derived from 
$M_m$ by replacing I by H and H by HR; the matrix $N_m$ models local operation 
patterns for the $m$-party entanglement assisted protocols. Separately, 
matrices $M_3$ and $M_4$ are used in similar fashion in 
Sections \ref{secfoddevenparity} and \ref{secmultihamming}.
Although the two categories of problems and their protocols differed, one 
yielding to entanglement assistance and the other succumbing to classical 
means with no entanglement whatsoever, the unifying aspect was the common
or similar pattern of local operations. Local operations are compactly 
represented by the elegant recursively defined matrices
$M_m$ and $M'_m$. These matrices are based
on the four element group represented by the 
matrix $V_4$ (see Section \ref{secfoddevenparity}). 
To the best of our knowledge, the matrices $M_m$ and
$M'_m$ do not appear in the literature. We feel that these 
matrices or similar recursively defined 
structures may be useful in compactly representing local operations for 
quantum entanglement assisted protocols for other classes of problems too.

Before concluding, we also consider the two-party 
scenario for mixed parity promise restricted functions. 
In contrast to the celebrated
linear lower bound on the deterministic classical communication complexity  
of the two-party INNER PRODUCT function (see \cite{kre:qc,kn:cc}), 
the following function $g_{11}$ with mixed parity promise has a one cbit 
classical protocol. 
Following uniform notation, we define $g_{11}(X,Y)=
\bigoplus_{i=1}^{n} x_i \wedge y_i$, where $x_iy_i$ is restricted
to be from the mixed parity promise set $\{11,01,10\}$. We construct a
deterministic classical one-cbit protocol where Alice 
and Bob first come up with 
bits $a$ and $b$, respectively, so that $x_i\wedge y_i=a\oplus b$.  
Let $a$ be called $a_0$ if $x_i=0$, and
$a_1$, otherwise. 
Likewise, let $b$ be called $b_0$ if $x_i=0$, and
$b_1$, otherwise. 
We must have (i) $a_1\oplus b_1=1$, (ii) $a_o\oplus b_1=0$, and 
(iii) $a_1\oplus b_0=0$. Note that assigning 1 to $a_1$ and $b_0$, and 
0 to $a_0$ and $b_1$  
achieves our purpose, leading to a 1-cbit classical protocol for Alice 
coming up with the value of $g_{11}(X,Y)$. Indeed, we assert that 
the result analogous to
Theorem \ref{theoremgAm} holds also for the two-party case.

We have studied only one-round, constant communication complexity
protocols. We propose that problems yielding to multiple rounds 
be investigated and characterized. We believe that such low 
communication complexity problems for various input promise sets would be 
very useful in VLSI design and also in mobile distributed computing.
We have presented results pertaining only to deterministic computations. A
natural research direction is the study
of probabilistic computations of partial boolean 
functions requiring constant or low communication complexity. 
Another important problem is that of settling the optimal 
classical communication complexity bound for the $m$-party (partial)
functions in Section \ref{secmultihamming}.
 
\vspace{.1in}

\noindent{\bf Acknowledgements:} The authors thank Siddhartha 
Brahma of Princeton University for valuable discussions, and Lov Grover of 
Bell Labs., Lucent, and R. Srikant of RRI, Bangalore for encouragement and 
advice. S. P. Pal and S. Kumar thank Guruprasad Kar and 
Samir Kunkri of ISI Kolkata for useful 
insights on contextuality and non-locality in quantum mechanics.
\bibliographystyle{plain}

\end{document}